%% file: main.tex
\documentclass[11pt]{article}
\usepackage[letterpaper,margin=1in]{geometry}
\usepackage{amsmath,amssymb,amsthm,mathtools}
\usepackage{microtype,booktabs,tabularx,xcolor}
\usepackage{enumitem}
\usepackage{algorithm}
\usepackage[noend]{algpseudocode}
\usepackage{tikz}
\usetikzlibrary{arrows.meta,positioning,calc,decorations.pathreplacing}
\usepackage[colorlinks=true,linkcolor=blue!55!black,citecolor=blue!55!black,urlcolor=blue!55!black]{hyperref}
\numberwithin{equation}{section}
\setlist[itemize]{leftmargin=1.5em,itemsep=3pt,parsep=0pt,topsep=5pt}
\setlist[enumerate]{leftmargin=1.7em,itemsep=3pt,parsep=0pt,topsep=5pt}

\newtheorem{theorem}{Theorem}[section]
\newtheorem{lemma}[theorem]{Lemma}
\newtheorem{proposition}[theorem]{Proposition}
\newtheorem{corollary}[theorem]{Corollary}
\theoremstyle{definition}

\newcommand{\ALG}{\operatorname{ALG}}
\newcommand{\OPT}{\operatorname{OPT}}
\newcommand{\E}{\mathbb{E}}

\newcommand{\rank}{\operatorname{rank}}
\newcommand{\Pending}{\mathcal P}
\newcommand{\Marked}{\mathcal G}
\definecolor{PaperBlue}{HTML}{315B88}
\definecolor{PaperTeal}{HTML}{21776F}
\definecolor{PaperAmber}{HTML}{A46422}
\hypersetup{linkcolor=PaperBlue,citecolor=PaperTeal,urlcolor=PaperBlue,
  pdftitle={List Update with Per-Page Maximum Delay},
  pdfauthor={Tianhang Lu, Runtian Ren, Shengcai Liu}}
\title{List Update with Per-Page Maximum Delay}
\author{Tianhang Lu\qquad Runtian Ren\qquad Shengcai Liu\\[0.65em]
{\small Guangdong Provincial Key Laboratory of Brain-Inspired Intelligent Computation}\\
{\small Department of Computer Science and Engineering}\\
{\small Southern University of Science and Technology, Shenzhen 518055, China}\\
{\small \texttt{liusc3@sustech.edu.cn}}\\[0.3em]
}
\date{}

\begin{document}
\maketitle
\thispagestyle{empty}
\input{sections/abstract}
\input{sections/introduction}
\input{sections/model}
\input{sections/algorithm_overview}
\input{sections/comparison_overview}
\input{sections/offline_overview}
\input{sections/extensions_overview}
\input{sections/limits_overview}
\clearpage
\input{sections/technical_online}

\input{sections/technical_offline}
\input{sections/technical_extensions}
\input{sections/technical_lower}
\input{sections/disclosure}
\bibliographystyle{plain}
\bibliography{references}
\end{document}

%% file: sections/abstract.tex
\begin{abstract}
We study list update with \emph{per-page maximum delay}. Requests may wait,
and one access clears all pending requests of a page at its current list
rank, paying their maximum waiting penalty. Each service can change the
permutation and therefore the cost of later accesses. We give a strictly
$5$-competitive deterministic algorithm: serve a page when the oldest
waiting time reaches two thirds of its current rank, then move it to the
front.

Our main tool is an \emph{asynchronous marked-rank coupling} that compares
different list orders and different service histories. A temporal transfer
principle separates this permutation argument from two waiting-cost
inequalities. It gives factor $5$ for page-specific superadditive penalties,
common calendar rates, and additional hard deadlines; deadlines alone
admit factor $3$. Two structural conditions on request-specific penalties
give factor $3+2\kappa$. In particular, nonincreasing request severity
retains factor $5$ even with unbounded weight ratios.

For linear delay, offline optimization is NP-complete but fixed-parameter
tractable in the number of pages. We characterize fixed-list batching,
show an $\Omega(m)$ gap between the best static order and dynamic
reordering, and prove that the optimal deterministic factor for one page
is exactly $2$. Counterexamples delineate the threshold rule's scope.
\end{abstract}

%% file: sections/introduction.tex
\section{Introduction}\label{sec:introduction}
A list provides a simple way to store objects that are accessed repeatedly.
Finding an object near the front is inexpensive, while finding an object
near the end requires a longer scan. Moving accessed objects forward can
reduce later search costs. Classical list update studies this reordering
decision when each request must be served in sequence. Move-to-front
already gives a constant competitive guarantee in that setting
\cite{sleator1985}.

When service may be postponed, reordering is only part of the decision.
Several requests for one page can be cleared by a single access. Waiting
can avoid repeated searches, but it also leaves an outstanding obligation.
Under a request-additive objective, each new request adds its own waiting
cost. We consider a different measure: an access pays the maximum waiting
time among the requests it clears. Younger requests for the same page
share its current waiting charge. Requests for different pages require
different accesses and pay separate charges.

We call this problem \emph{list update with per-page maximum delay}.
The algorithm decides when to access each pending page and how to change
the list. Accessing a page costs its current rank, counted from one, plus
the age of its oldest pending request. The page may then move toward the
front for free. Paid adjacent exchanges cost one and are permitted at any
time. We compare with an offline optimum that knows the complete timed
input and may choose all service times and list operations.

The persistent list makes the timing decision consequential. Serving a
page and moving it forward can make another pending page more expensive
to access. Conversely, a page that the online algorithm keeps waiting may
already have been served cheaply by the offline optimum. Its waiting
cannot then be charged to simultaneous offline delay. A successful
analysis must account for both different list orders and different sets
of completed requests.

\subsection{Our contributions}\label{sec:contrib}
Our results concern \emph{one access per page}, where repeated requests
to that page share both the access and the per-batch maximum-delay charge.
Classical list update has a changing permutation but no scheduling,
whereas maximum-delay aggregation with fixed service costs has scheduling
but no persistent list state.
\begin{itemize}
\item \textbf{A constant guarantee from asynchronous coupling.}
We give a simple, deterministic, strictly $5$-competitive algorithm,
independently of the list length and against an unrestricted clairvoyant
comparator. The key is a marked-rank inequality for \emph{any} completing
move-to-front (MTF) service schedule. A temporal transfer proposition
then turns this spatial comparison into a competitive guarantee.
\item \textbf{Delay and deadline extensions.}
The same proof interface gives factor $5$ for page-specific
superadditive waiting penalties, common calendar rates, and additional
hard deadlines. Deadlines without waiting penalties admit factor $3$.
Two distinct conditions on request-specific penalties yield $3+2\kappa$:
a bounded superadditive page baseline, or pointwise domination of
later requests by $\kappa$ times every earlier request on the same page.
The latter allows arbitrarily large weight ratios when severity is
nonincreasing, with factor $5$.
\item \textbf{Offline structure and sharp boundaries.}
For linear waiting, exact offline optimization is NP-complete but
fixed-parameter tractable in the number of pages. We characterize
fixed-list batching and show that even the best freely chosen static
order can lose $\Omega(m)$ against a dynamic schedule. For one page
the optimal deterministic competitive ratio is exactly $2$.
The proposed multi-page threshold rule has a lower-bound family
approaching $10/3$, and counterexamples explain where its delay
assumptions matter.
\end{itemize}

\subsection{Our techniques: separating order from time}\label{sec:techniques-intro}
\paragraph{Why ordinary inversions are not enough.}
The classic MTF inversion potential measures disagreement between two
permutations. Here the schedules may also disagree about \emph{which
requests have already been served}. In particular, the comparator may
have cleared a request while it is still pending online; charging all
subsequent online waiting to simultaneous offline waiting would fail.
We mark such an online page episode when the comparator serves its
oldest request and retain the mark until the online access.
For the online and comparator permutations $A$ and $O$, use
\begin{equation}\label{eq:intro-potential}
 \Phi=I(A,O)+2\sum_{p\text{ marked}}r_O(p).
\end{equation}
The extra credit is the \emph{offline rank}, not merely one token per
marked page. An MTF access crosses the pages before its target; the
rank-dependent credit is exactly what makes these pairwise inversion
changes affordable. Checking accesses and paid exchanges yields the
purely spatial inequality
\begin{equation}\label{eq:intro-coupling}
 C_A\le 3C_O+2B_A,
\end{equation}
where $B_A$ sums the access ranks of online episodes that are still
unmarked when served. No waiting rule is used in this comparison.

\paragraph{Why the remaining accesses can be charged to time.}
If an online episode ends unmarked, its oldest request has remained
pending for the comparator throughout that episode. For each page,
these online intervals are disjoint and lie inside comparator pending
intervals. A rank-aware threshold therefore supplies
$D_A\le\beta C_A$ and $\beta B_A\le\kappa D_O$ whenever the delay
model supports this interval charge. The temporal transfer principle
(Proposition~\ref{prop:temporal-transfer}) combines these conditions
with~\eqref{eq:intro-coupling}, yielding a strict factor
$3+2\kappa$ at $\beta=2\kappa/3$. Ordinary linear delay has
$\kappa=1$, hence factor $5$.

\paragraph{Why the extensions share one proof.}
A hard deadline can force an earlier service, but that service is
already marked because the comparator must meet the same deadline.
For nonlinear or heterogeneous penalties, the permutation argument
stays fixed; only the justification of the unmarked-interval charge
changes. Superadditivity, a common calendar rate, a bounded page
baseline, or domination by an earlier request's function each supplies
a different way to pay for those intervals. The same separation also
makes the scope of the theorem transparent: shared-prefix scans or
set-valued requests would need a different spatial comparison.

\paragraph{Why the offline state compresses.}
Under linear per-page delay, holding cost is the number of pending
pages. After charging elapsed waiting, an offline dynamic program
needs only the current permutation and pending-page set; widely spaced
arrivals recover the hardness of classical offline list update.

\subsection{What the guarantees compare}
Let $m$ be the number of pages and $N$ the number of request occurrences.
All online bounds compare complete costs on each finite input;
they have no additive term. The exact offline state space has size
$S=m!2^m$, so its dependence on input length is polynomial but its
dependence on $m$ is not. Table~\ref{tab:results} separates these guarantees
from bounds on a particular online rule.

\begin{table}[htbp]
\centering\small
\begin{tabularx}{\linewidth}{@{}p{.30\linewidth}X@{}}
\toprule
Setting & Result \\
\midrule
Linear waiting, online & Deterministic strict factor $5$; the chosen
threshold rule has a lower bound approaching $10/3$. \\
Page-wise superadditive penalties & Strict factor $5$; includes convex waiting and common exogenous calendar rates. \\
Hybrid with deadlines & Strict factor $5$ for common
page penalties; strict factor $3+2\kappa$ for request-specific penalties
within a known factor $\kappa$ of a superadditive page baseline. \\
Arrival-ordered penalties & Strict factor $3+2\kappa$ under one-sided function domination; factor $5$ when request severity is nonincreasing, even without bounded weight ratios. \\
Hard deadlines only & Strict factor $3$ with one access per pending page. \\
Linear waiting, offline & NP-complete decision problem; exact
$O(NSm\log S)$ arithmetic-operation algorithm, $S=m!2^m$; polynomial-time
$5$-approximation. \\
One page, linear waiting & Optimal deterministic factor exactly $2$; the lower bound applies even allowing an additive constant. \\
Fixed list & Explicit optimal batching in linear time after requests are
grouped by page. \\
Best free static order & Polynomial-time assignment formulation, but
$\Omega(m)$ loss against the dynamic optimum is possible. \\
\bottomrule
\end{tabularx}
\caption{The algorithm-specific lower bound does not apply to every online
algorithm. The nonlinear theorem does not imply the stated offline state
compression beyond linear waiting.}\label{tab:results}
\end{table}

\subsection{Related work and model distinctions}\label{sec:related}

\paragraph{Classical list update and movement rules.}
The classical list-update problem serves each requested item immediately,
charging its current position and the cost of rearranging the list.
Move-to-front is $2$-competitive in the standard model, which allows a
requested item to move toward the front for free after an access
\cite{sleator1985}; randomized list-update rules can further improve the
competitive ratio \cite{albers1995}.
Offline list update is NP-hard \cite{ambuhl2000}, and this hardness
carries over to our model by spacing requests sufficiently far apart.
Recent work by Basiak, Bienkowski, B\"ohm, Chrobak, Je\.{z}, Sgall, and
Tatarczuk gives a $3.3904$-competitive algorithm for \emph{list update
with uniform costs}, where \emph{every} adjacent exchange must be paid
for \cite{basiak2025uniform}.  This differs from the standard free-forward
model used here.  In particular, its numerical competitive ratio is not
a bound for our delayed problem, and the values should not be directly
compared.

\paragraph{Set-valued requests and dynamic rankings.}
Online min-sum set cover (MSSC) generalizes instantaneous list update:
each request specifies a set $R$ of acceptable items, and its access cost
is the rank of the earliest item of $R$ in the maintained permutation.
Fotakis, Kavouras, Koumoutsos, Skoulakis, and Vardas initiated the
competitive study of this online problem, including comparisons with
static and dynamic permutations \cite{fotakis2020mssc}.
For sets of size at most $r$, Bienkowski and Mucha obtained a randomized
$O(r^2)$-competitive algorithm against the \emph{dynamic} optimum, as
well as an existential deterministic $O(r^4)$ guarantee
\cite{bienkowski2023mssc}.  Basiak, Bienkowski, and Tatarczuk subsequently
gave a constructive deterministic $O(r^2)$-competitive algorithm against
the dynamic optimum \cite{basiak2023mssc}.
These results are closely related through the persistent permutation and
the dynamic benchmark, but their requests are served upon arrival and may
be satisfied by \emph{different} chosen items in the online and offline
schedules.  Our requests instead name a unique page, and the new decision
is \emph{when} to clear its pending requests.  Thus MSSC generalizes the
spatial request type of classical list update, not the delayed batching
model studied here.

\paragraph{List update with time windows and additive delays.}
The closest prior work on delaying list accesses is by Azar, Lewkowicz,
and Vainstein \cite{azar2024list}.  They study time windows and
request-additive delay functions, obtaining constant competitive
algorithms; the published results include factors $24$ for time windows
and $336$ for general delays.  Their service operation can scan a
prefix of the list once, satisfying requests for \emph{several different
pages} at the cost of the farthest position reached.
Our operation accesses \emph{one} page, clears all requests currently
pending for that page, and charges one maximum delay for that page's
oldest pending request.  Distinct pages pay separate access and delay
charges, even when served at the same instant.  Replacing one prefix
scan by a sequence of per-page accesses may change a spatial cost of
$k$ into $1+\cdots+k$, while replacing request-additive penalties by a
per-page maximum changes the temporal accounting.  Neither algorithmic
guarantee therefore directly transfers between the two models, nor does
a comparison of their constants alone indicate an improvement.

\paragraph{Predictions and maximum-delay service systems.}
A different information-based variant, list update with predictions,
was studied by Azar, Lewkowicz, and Suriyanarayana
\cite{azar2025prediction}; it addresses accuracy--robustness tradeoffs
for predicted accesses rather than waiting and batching.
On the temporal side, maximum-delay aggregation under fixed service
costs has constant-competitive algorithms
\cite{lu2026maximum,lu2026covering}.  Those service systems do not
retain a changing list permutation after each action.  Our marked-rank
comparison addresses the additional difficulty that online and offline
schedules may have \emph{both} different list orders and different sets
of requests already served.  The resulting coupling is specific to
per-page episodes; extending it to prefix-wide or set-valued services
would require additional accounting for shared access costs or for
requests served through different elements.

\paragraph{Organization.}
The first part develops the model, algorithm, asynchronous comparison,
offline structure, and the common principle behind the extensions. Complete event proofs,
offline recurrences, reductions, and lower-bound constructions follow
in Sections~\ref{sec:proof-online}--\ref{sec:proof-lower}.

%% file: sections/model.tex
\section{The Model and Its Waiting Cost}\label{sec:model}
An input consists of a finite sequence of requests $(p_i,a_i)$, where
$p_i$ is one of $m$ pages and $a_i\ge0$ is its arrival time. Requests with
the same arrival time are revealed together, before any actions at that
time. The initial permutation is specified and shared by the algorithm
and comparator. The final permutation is unrestricted.

\paragraph{Access and rearrangement.}
At any time a schedule may access a page with pending requests. If its
current rank is $r$, the access costs $r$ and clears every pending request
for that page. The schedule may immediately move that page to any earlier
position for free, including leaving it where it is. Other changes are
implemented by adjacent exchanges at unit cost. These exchanges may be
performed proactively. Operations take no time; simultaneous accesses
are ordered, and their ranks are evaluated in that order.

Passing other pages while locating the target does not serve their
requests. Thus requests to one page can share an access, whereas
requests to different pages cannot share one prefix scan's cost.

\paragraph{Pending episodes.}
When a request arrives at a page with no pending request, it starts an
\emph{episode}. Write $\alpha$ for this first arrival. Further requests
to the page join the episode without changing $\alpha$. If the next
access occurs at $t$, the episode contributes delay $t-\alpha$. Thus a
schedule has total cost
\begin{equation}\label{eq:model-cost}
 C+D
 =\sum_{\text{accesses}}\rank(p)
  +\#\{\text{paid exchanges}\}
  +\sum_{\text{episodes }[\alpha,t]}(t-\alpha).
\end{equation}
The algorithm must eventually clear every request on every finite input,
without being told that no more requests will arrive. Its cost includes
all accesses and waiting after the final arrival.

\begin{proposition}[One clock per pending page]\label{prop:holding-story}
If $\Pending(t)$ is the set of pending pages in a schedule, then
\begin{equation}\label{eq:holding}
 D=\int_0^\infty |\Pending(t)|\,dt.
\end{equation}
\end{proposition}
Each page contributes one while an episode is active and zero otherwise.
Integrating gives exactly its episode lengths. This identity is both an
online accounting rule and the reason that the offline dynamic program
can forget ages after charging past waiting.

\paragraph{Two distinctions in the objective.}
If ten requests arrive at one pending page, its holding rate is still
one. If ten different pages are pending, the rate is ten. Consequently,
the objective is neither the sum of individual request delays nor one
maximum taken over the entire execution. Even when several pages are
accessed at the same time, each page pays its own access and episode
delay. We do not combine their ranks or waiting costs into one pooled
service charge.

\paragraph{Benchmarks and computation.}
The offline optimum may use every permitted list operation and choose
service times with full knowledge of the input. A strict factor $R$
means $\ALG(I)\le R\OPT(I)$ for every finite timed input $I$. We use
rational arrival times for bit-complexity and decision statements. For
general continuous penalty functions, the online guarantee is stated
with access to their current values and first threshold-hitting times;
it does not assert polynomial bit complexity for arbitrary function
representations.

%% file: sections/algorithm_overview.tex
\section{Choosing When to Access}\label{sec:algorithm-story}
The algorithm compares an episode's age with the cost of accessing its
page now. A page near the front should wait less than a page near the end.
This comparison must remain current: the algorithm does not freeze the
page's rank when its episode begins.

Fix $\beta>0$. The \emph{scaled rank-threshold rule} accesses a pending
page $p$ as soon as
\begin{equation}\label{eq:trigger}
 t-\alpha_p\ge\beta r_A(p),
\end{equation}
then moves $p$ to the front. It recomputes all eligibility conditions after
every access. Ties may be resolved arbitrarily. The algorithm performs no
paid exchanges.

\begin{algorithm}[htbp]
\caption{Scaled Rank-Threshold MTF}\label{alg:threshold}
\begin{algorithmic}[1]
\State Maintain the list and the oldest arrival $\alpha_p$ of each pending page.
\State When a request arrives at a nonpending page $p$, set $\alpha_p$ to its arrival time.
\State Reveal all arrivals at the current time before taking an action.
\While{some pending page satisfies $t-\alpha_p\ge\beta r_A(p)$}
  \State Choose one such page $p$, access it, and clear its pending requests.
  \State Move $p$ to the front and remove its episode clock.
\EndWhile
\State Wait for an arrival or the next equality in~\eqref{eq:trigger}, whichever occurs first.
\end{algorithmic}
\end{algorithm}

\paragraph{Why every service hits its threshold exactly.}
While $p$ is pending, another page's move-to-front can only preserve or
increase $p$'s rank. It cannot suddenly lower the threshold below the
current age. Between actions, age increases continuously. Thus every
actual access satisfies equality in~\eqref{eq:trigger}, even when an
earlier action at the same time postpones another page's service.
Writing $C_A$ for the sum of access ranks, we obtain
\begin{equation}\label{eq:balance-story}
 D_A=\beta C_A,\qquad \ALG=(1+\beta)C_A.
\end{equation}
Since ranks never exceed $m$, each episode completes within $\beta m$
time units of its first arrival. The rule therefore handles the end of
the input through its ordinary timers.

\begin{theorem}[Scaled threshold guarantee]\label{thm:main}
For every $\beta>0$ and every feasible comparison schedule with access
and paid-exchange cost $C_O$ and delay $D_O$,
\begin{equation}\label{eq:main-comparison}
 \ALG_\beta\le3(1+\beta)C_O+
             \frac{2(1+\beta)}{\beta}D_O.
\end{equation}
In particular, $\beta=2/3$ gives a strict deterministic factor $5$.
\end{theorem}
The two coefficients measure different losses. A larger threshold increases
the waiting attached to each access, while a smaller threshold makes it
harder to charge early online services to offline waiting. Their maximum
is minimized at $\beta=2/3$. The theorem does not assert that this choice
or the resulting factor is optimal among all algorithms.

\subsection{A service can postpone another page's timer}\label{sec:worked}
Consider the initial list $[A,B,C]$ and $\beta=2/3$. Requests for $B$ and
$C$ arrive at time zero. Another request for $B$ arrives at time $1/3$,
and a request for $A$ arrives at time one. The second $B$ request joins
the existing episode; it does not restart its clock or add a second
waiting charge.

At time $4/3$, page $B$ has age $4/3$ and rank two, so it is accessed and
moved to the front. The list becomes $[B,A,C]$. Page $A$ had been due at
$1+2/3=5/3$, but its rank is now two and its threshold moves to $7/3$.
At time two, $C$ reaches its rank-three threshold. Moving it forward
changes the list to $[C,B,A]$ and raises $A$'s rank again. Page $A$ is
finally accessed at time three, with age two and rank three.

\begin{figure}[htbp]
\centering
\begin{tikzpicture}[x=3.1cm,y=.65cm,>=Latex,font=\small]
\draw[->,gray] (-.08,-.75)--(3.35,-.75) node[right]{$t$};
\foreach \x/\lab in {0/0,1/1,2/2,3/3}{
 \draw[gray] (\x,-.81)--(\x,-.69);
 \node[below] at (\x,-.81){$\lab$};}
\node[left] at (-.1,0){$A$};
\node[left] at (-.1,1){$B$};
\node[left] at (-.1,2){$C$};
\draw[very thick,PaperBlue] (1,0)--(3,0);
\draw[very thick,PaperTeal] (0,1)--(1.3333,1);
\draw[very thick,PaperAmber] (0,2)--(2,2);
\foreach \p in {(1,0),(0,1),(.3333,1),(0,2)}{\fill \p circle (2pt);}
\foreach \p in {(3,0),(1.3333,1),(2,2)}{
 \draw[fill=white,thick] \p circle (3pt);}
\node[above,PaperTeal] at (1.3333,1.05){access $B$};
\node[above,PaperAmber] at (2,2.05){access $C$};
\node[above,PaperBlue] at (3,.05){access $A$};
\end{tikzpicture}
\caption{Filled dots are arrivals; open circles are accesses. An episode's
horizontal length is its single maximum-delay charge. The two $B$ arrivals
share one episode.}\label{fig:timers}
\end{figure}
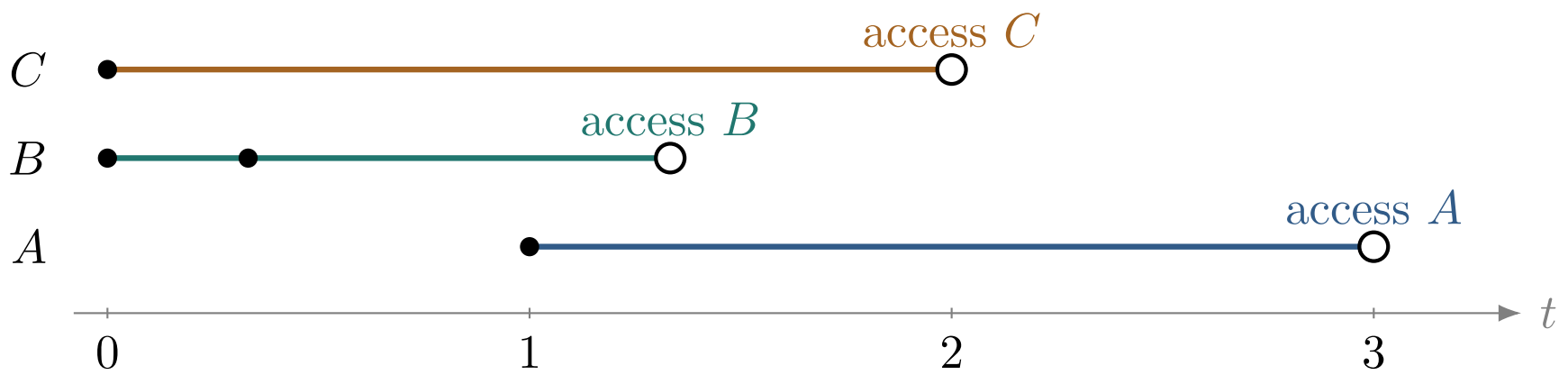

\begin{table}[htbp]
\centering\small
\begin{tabular}{@{}ccccc@{}}
\toprule
Access time & Page & Rank & Episode age & List after access \\
\midrule
$4/3$ & $B$ & $2$ & $4/3$ & $[B,A,C]$ \\
$2$ & $C$ & $3$ & $2$ & $[C,B,A]$ \\
$3$ & $A$ & $3$ & $2$ & $[A,C,B]$ \\
\bottomrule
\end{tabular}
\end{table}

The access total is eight, and delay is $16/3$, exactly two thirds of
that total. A comparator could keep the original list, access $B$ at
$1/3$, $C$ at zero, and $A$ at one, paying $6+1/3$ in total. Its batches
and list movements need not match those of the algorithm. The proof
therefore cannot pair service events merely by their chronological order.

The example also explains why an initial-rank timer is a different
algorithm. Freezing $A$'s original rank would serve it at $5/3$ after its
access cost had changed. Our rule keeps the exact relation between the
cost of a service and the waiting accumulated for it.

%% file: sections/comparison_overview.tex
\section{Comparing Different Service Times}\label{sec:comparison-story}
The main difficulty is the part of online waiting that takes place after
the comparator has already served the same request. We distinguish this
case explicitly rather than charging every online clock to an offline
clock running at the same time.

\subsection{Which episodes can pay directly to offline waiting?}
During an online episode for page $p$, mark that page when the comparison
schedule first serves the episode's oldest request. The mark remains
until the online algorithm accesses $p$. Later arrivals to $p$ do not
change the identity of that oldest request, and do not remove the mark.
These marks are used only in the proof; the online algorithm never sees
the comparator.

Suppose an online episode ends while its page is still unmarked. Its
oldest request has been pending offline throughout that episode. The
whole online interval can therefore be placed inside one offline pending
interval of the same page. Distinct online episodes for a page have
disjoint interiors. If $B_A$ is the sum of access ranks over these
unmarked services, threshold equality gives
\begin{equation}\label{eq:bad-story}
 \beta B_A\le D_O.
\end{equation}
This charge does not count requests separately. An offline episode may
contain many online episodes, but the sum of their lengths is at most
the offline interval's length.

\begin{figure}[htbp]
\centering
\begin{tikzpicture}[x=.95cm,y=.85cm,>=Latex,font=\small]
\node[left] at (0,1){Offline $p$};
\node[left] at (0,0){Online $p$};
\draw[very thick,PaperAmber] (.3,1)--(9,1);
\draw[PaperAmber] (.3,.87)--(.3,1.13) (9,.87)--(9,1.13);
\draw[very thick,PaperBlue] (.9,0)--(2.4,0);
\draw[very thick,PaperBlue] (3.0,0)--(5.4,0);
\draw[very thick,PaperBlue] (6.2,0)--(8.2,0);
\foreach \x in {.9,3,6.2}{\fill (\x,0) circle (2pt);}
\foreach \x in {2.4,5.4,8.2}{\draw[fill=white] (\x,0) circle (3pt);}
\draw[densely dashed,gray] (.9,.1)--(.9,.9) (8.2,.1)--(8.2,.9);
\node[above] at (4.65,1.1){one offline episode};
\node[below] at (4.65,-.15){disjoint unmarked online episodes};
\end{tikzpicture}
\caption{For a fixed page, unmarked online intervals can be charged to
offline pending time without overlap. Marks are needed for the other
online intervals, which may continue after offline service.}
\label{fig:intervals}
\end{figure}
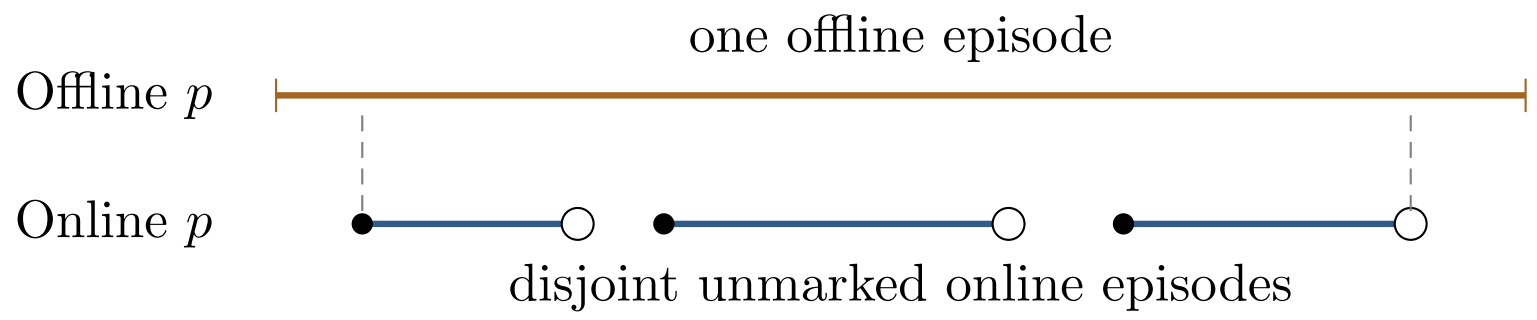

\subsection{What must be remembered after offline service?}
If the comparator serves a request early, its online episode may still
be expensive because the online page is far back in the list. The
comparison keeps credit for this event at the page's \emph{offline}
rank. That credit changes when the comparator rearranges its list and
is released when the online algorithm finally clears the episode.

The rank is important. The number of marked pages alone does not record
how many pairwise order differences must be removed by a later
move-to-front. The offline rank supplies precisely the quantity that
appears in the inversion calculation on the next page.

\subsection{One potential for order and unfinished episodes}
Let $I(A,O)$ count unordered page pairs whose relative order differs in
the two lists. If $\Marked$ is the set of marked pages, define
\begin{equation}\label{eq:potential-story}
 \Phi=I(A,O)+2\sum_{p\in\Marked}r_O(p).
\end{equation}
It starts at zero and is always nonnegative. Arrivals and waiting do not
change it. Only accesses, free forward moves, paid exchanges, and the
associated mark changes need to be checked.

\paragraph{An online access removes order disagreements.}
Suppose the algorithm accesses $p$ at rank $k$, while its offline rank is
$\ell$. Let $a$ be the number of pages preceding $p$ in both lists.
Moving $p$ to the front changes the inversion count by
\[
 \Delta I=2a-(k-1).
\]
For a marked page, clearing its episode also removes $2\ell$ from the
potential. Since $a\le\ell-1$, this decrease pays for the access:
$k+\Delta\Phi\le-1$. For an unmarked page, no credit is removed, but
$\Delta I\le k-1$ gives $k+\Delta\Phi\le2k$. These are exactly the
services already covered by~\eqref{eq:bad-story}.

\paragraph{Offline changes pay for the credit they create.}
An offline access at rank $\ell$ may create a mark of value $2\ell$.
Moving that marked page forward reduces its own rank contribution and
increases the ranks of at most the pages it passes. The marked-rank sum
cannot increase during this move. At most $\ell-1$ new inversions are
created. If no mark is created, the same crossed pages bound the increase
directly. Each paid adjacent exchange changes at most one inversion and
two units of marked-rank credit.

\begin{table}[htbp]
\centering\small
\begin{tabularx}{\linewidth}{@{}Xl@{}}
\toprule
Event & Bound used in the sum \\
\midrule
Online access, marked page, rank $k$ & $k+\Delta\Phi\le0$ \\
Online access, unmarked page, rank $k$ & $k+\Delta\Phi\le2k$ \\
Offline access and free forward move, rank $\ell$ & $\Delta\Phi\le3\ell$ \\
Offline paid adjacent exchange & $\Delta\Phi\le3$ \\
\bottomrule
\end{tabularx}
\caption{The four event types behind the marked-rank comparison.
Complete case calculations appear in Section~\ref{sec:proof-online}.}
\label{tab:marked-rank-events}
\end{table}

Summing and discarding the nonnegative final potential yields the
combinatorial comparison
\begin{equation}\label{eq:coupling-story}
 C_A\le3C_O+2B_A.
\end{equation}
It does not use the threshold rule. What remains is a temporal
certificate for the unmarked accesses; the list comparison is reused
without change in every extension below.

\subsection{A transfer from temporal charges to total cost}
The next proposition states the reusable interface between the
permutation argument and a waiting policy. It does not assume a specific
delay function or a threshold-equality identity.
\begin{proposition}[Temporal transfer]\label{prop:temporal-transfer}
Let $A$ be a feasible, completing MTF algorithm that makes no paid
exchanges and clears every pending request at its accessed page.
Fix a feasible comparator $O$ with the same initial list, and let $B_A$
be the sum of ranks of $A$'s accesses unmarked relative to $O$.
Suppose that for some $\beta>0$ and $\kappa>0$,
\begin{equation}\label{eq:transfer-assumptions}
 D_A\le\beta C_A,\qquad \beta B_A\le\kappa D_O.
\end{equation}
Then
\begin{equation}\label{eq:transfer-bound}
 \ALG\le3(1+\beta)C_O+\frac{2\kappa(1+\beta)}{\beta}D_O.
\end{equation}
If these conditions hold on every finite input against every feasible
comparator for a fixed $\kappa>0$ and $\beta=2\kappa/3$, then $A$ is
strictly $(3+2\kappa)$-competitive. For the fixed comparator $O$,
$D_A=B_A=0$ instead gives $C_A\le3C_O$. If there are no waiting
penalties and every access is marked on every finite input against
every feasible comparator, then $A$ is strictly $3$-competitive.
\end{proposition}
The proof in Section~\ref{sec:proof-online} combines the coupling
with~\eqref{eq:transfer-assumptions}. For linear waiting, both conditions
hold with $\kappa=1$ and threshold equality gives $D_A=\beta C_A$.
With deadlines, an early deadline-driven access is already marked,
so equality for all online services is replaced by $D_A\le\beta C_A$;
the unmarked-access charging inequality is still required.

%% file: sections/offline_overview.tex
\section{Offline Structure}\label{sec:offline-story}
The online algorithm gives a polynomial-time $5$-approximation offline,
but exact optimization reveals additional structure. The linear waiting
objective permits a finite state description that does not retain the
arrival time of each pending episode.

\paragraph{Charge waiting before forgetting its age.}
Between two consecutive arrival epochs, move every operation to the left
endpoint while preserving its order. No request arrives in the open gap,
the final permutation and pending set are unchanged, and any episode
cleared earlier has smaller delay. An optimum can therefore act only at
arrival epochs, including the final completion actions.

At a layer, store $(\pi,P)$: the current permutation and the pending-page
set. Across a gap $\Delta$, add $\Delta|P|$. All past waiting is now paid,
so the remaining problem does not depend on how long each page has
already waited. New arrivals take the union with $P$. Within the layer,
paid exchanges change $\pi$, while an access pays its rank, removes one
page from $P$, and permits every free forward destination. Shortest-path
closure in this finite graph includes arbitrary orders of same-time
actions.

\begin{theorem}[Offline computation]\label{thm:offline-story}
For linear delay and $S=m!2^m$, the offline optimum can be computed in
$O(NSm\log S)$ arithmetic operations and $O(Sm)$ space. With rational
input, the decision problem is NP-complete.
\end{theorem}
For this bound, an access followed by a free forward move is represented
by paying for the move before the access and then paying its lower rank;
the total is unchanged. The state bound gives fixed-parameter tractability in $m$, rather than an
exponent of $m$ on the input length. The NP-hardness reduction places
classical list-update requests so far apart that optimal waiting cannot
cross the next arrival. Membership in NP follows by shortening paid-swap
paths inside each arrival layer. Section~\ref{sec:proof-offline} supplies
both arguments and the bit-complexity qualifications.

\paragraph{The option not to move matters.}
In the initial list $[A,B]$, let $B$ be requested at zero and $A$ at time
ten. Accessing $B$ immediately and leaving it in place, then accessing
$A$, costs three. Forcing move-to-front after every offline access changes
this instance's optimum to four. An exact oracle must therefore include
all optional forward destinations, not just move-to-front.

\paragraph{A fixed list is simpler, but insufficient.}
If a page has fixed rank $c$ and distinct arrivals $t_1<\cdots<t_q$, its
optimal cost is
\begin{equation}\label{eq:static-story}
 c+\sum_{i=1}^{q-1}\min\{t_{i+1}-t_i,c\}.
\end{equation}
Each gap can either remain inside a batch, paying its length, or separate
two batches, paying another access. Empty pages contribute zero. If the
initial static order may be chosen for free, assigning each page to a
distinct rank with cost~\eqref{eq:static-story} is a minimum-cost matching
problem. Nevertheless, even this best static order can be
$\Omega(m)$ more expensive than a dynamic schedule: long successive
phases requesting one page at a time reward moving the currently active
page forward. Exact static optimization cannot replace dynamic reordering.

%% file: sections/extensions_overview.tex
\section{Beyond Linear Waiting}\label{sec:extensions-story}
The inversion calculation does not refer to elapsed time. It needs only
to distinguish marked from unmarked services. This allows a broader
delay theorem whenever the costs of disjoint online waiting intervals
can be paid by the offline episode containing them.

For each page $p$, fix a function $f_p:[0,\infty)\to[0,\infty)$ that is
continuous, nondecreasing, unbounded, and zero at zero. All requests of
that page use the same function. An episode of duration $x$ pays
$f_p(x)$, equivalently the maximum of the individual $f_p$-penalties of
its requests. Assume further that
\begin{equation}\label{eq:superadditive-story}
 f_p(x+y)\ge f_p(x)+f_p(y)\qquad(x,y\ge0).
\end{equation}
This includes positive page-specific linear weights and nondecreasing
unbounded convex functions vanishing at zero. Different pages may use
different functions, such as $f_p(t)=w_pt^{q_p}$ with $q_p\ge1$.

\begin{theorem}[Page-specific waiting penalties]\label{thm:extension-story}
Access page $p$ when $f_p(t-\alpha_p)$ reaches
$\tfrac23 r_A(p)$, then move it to the front. Under the preceding
assumptions, this algorithm is strictly $5$-competitive.
\end{theorem}

\paragraph{What changes in the proof.}
Exact threshold equality still gives $D_A=\tfrac23 C_A$. For the offline
charge, consider unmarked online intervals of lengths $x_1,\ldots,x_s$
inside one offline episode of page $p$, of length $L$. They are disjoint,
so $\sum_i x_i\le L$. Hence
\[
 \sum_i f_p(x_i)
 \le f_p\!\left(\sum_i x_i\right)
 \le f_p(L).
\]
This replaces the linear interval-length comparison. The list potential,
including its treatment of paid exchanges and optional free moves,
stays unchanged. Continuity supplies exact first hitting, and
unboundedness ensures completion without another request.

\paragraph{A common calendar rate.}
There is another useful form of the same interval argument. Let page $p$
have an exogenous nonnegative rate $w_p(t)$, and charge an episode
$[a,b]$ the amount $\int_a^b w_p(t)\,dt$. Rates may vary over absolute
time but are shared by the two schedules. If they are locally integrable
and every tail integral is infinite, the accumulated cost is a continuous
clock that eventually reaches any threshold. Costs on disjoint intervals
add directly, giving the same factor five. The algorithm needs the current
accumulated clock, not future values of the rate.

\paragraph{The scope is per page.}
The theorem does not allow an arbitrary new waiting function for each
request to the same page. In that model the oldest request need not
determine the maximum penalty, and the costs of disjoint online episodes
need not sum to at most the cost of an offline episode containing them.
The counterexamples on the next page make both this distinction and the
role of~\eqref{eq:superadditive-story} explicit. The offline pending-set
compression in Section~\ref{sec:offline-story} is a linear-delay result;
past ages cannot in general be forgotten for nonlinear penalties.

\paragraph{Adding hard deadlines and bounded request heterogeneity.}
The same marked-rank method also tolerates hard deadlines. Mark a page
whenever the comparator has served \emph{any} request in its current
online episode. An algorithm access forced by a deadline is then always
marked. With a common superadditive penalty function per page, the
usual threshold rule combined with the earliest-deadline trigger
remains strictly $5$-competitive; if only deadlines are charged as
hard feasibility constraints, serving at the earliest pending deadline
is strictly $3$-competitive.
More generally, if every request's penalty lies between a page-specific
superadditive baseline and $\kappa$ times that baseline, the combined
algorithm is strictly $(3+2\kappa)$-competitive for known $\kappa$.
The full statement and proof appear in
Theorem~\ref{thm:hybrid-extension} and Corollary~\ref{cor:deadline-only-3}.

\paragraph{A complementary extension: arrival-order domination.}
A common page baseline is not the only way to charge heterogeneous
requests. Suppose every request has a superadditive penalty function,
and a later request for the same page has a function at most $\kappa$
times that of \emph{any} earlier request, pointwise in waiting time.
There need not be a fixed page baseline: severity may decrease without
bound, and increases need only be controlled relative to earlier
requests. An unmarked online episode lies inside one offline episode;
its oldest offline request supplies a single superadditive function
that dominates the penalties of all the online requests up to factor
$\kappa$. The same argument now gives a strict $3+2\kappa$ factor.
In particular, for $\kappa=1$ the original factor $5$ holds even
with arbitrarily large decreases in weights. This contrasts with the
rapidly \emph{increasing} weights in the negative example of
Section~\ref{sec:limits-story}. The full proof is
Theorem~\ref{thm:arrival-ordered}.

%% file: sections/limits_overview.tex
\section{Bounds and Limits of the Rule}\label{sec:limits-story}
The factor five is an upper bound for a particular deterministic
algorithm. It is useful to separate its limitations from lower bounds
that apply to all online algorithms.

\subsection{A lower bound for the scaled threshold}
Start with the list $(1,\ldots,m)$ and repeatedly request
$m,m-1,\ldots,1$, separating consecutive requests by $H>\beta m$.
Every online episode finishes before the next arrival. Each requested
page is at rank $m$, waits $\beta m$, and costs $(1+\beta)m$.
After one block the list returns to its initial order. In comparison,
keeping the initial list fixed and serving immediately costs
$m(m+1)/2$ per block. Therefore
\begin{equation}\label{eq:reverse-story}
 \frac{\ALG_\beta}{\OPT}
 \ge\frac{2(1+\beta)m}{m+1}.
\end{equation}
Repeating blocks absorbs any fixed additive constant. This family
approaches $10/3$ for our choice $\beta=2/3$. It approaches four for
$\beta=1$, whose proved upper bound is six. Neither statement is a
lower bound of four on the problem itself.

\subsection{Two failures beyond the delay assumptions}
\paragraph{Concave waiting.}
For one page with $f(t)=t^\gamma$, $0<\gamma<1$, a fixed threshold
$f(\text{age})=\beta$ serves after time $\tau=\beta^{1/\gamma}$.
Release $N$ requests at intervals $2\tau$. The algorithm serves them
separately and pays $(1+\beta)N$. Offline may batch all of them at the
last arrival for
\[
 1+[2\tau(N-1)]^\gamma.
\]
The ratio grows without bound. Concavity makes one long offline episode
much cheaper than the sum of the shorter online episodes. The interval
charge used by the theorem cannot hold in this case.

\paragraph{Request-specific weights.}
Even linear functions can violate the required sharing when their
weights vary within a page. For fixed $\beta>0$, take one page and
requests $i=0,\ldots,N-1$ with
\[
 a_i=1-4^{-i},\qquad w_i=2\beta\,4^i,
\]
and charge each batch the maximum of $w_i(t-a_i)$ over its requests.
A rule that serves when this maximum reaches $\beta$ clears request
$i$ before the next one arrives, paying $(1+\beta)N$ in total. At time
one, however, every weighted age is exactly $2\beta$, so a single offline
batch costs $1+2\beta$. The construction is a family of finite inputs;
it does not require infinitely many arrivals in finite time.

Both examples disprove the corresponding fixed-threshold extension.
They do not rule out a constant guarantee for a different algorithm in
those models. Complete calculations appear in
Section~\ref{sec:proof-lower}.

\subsection{Problem lower bounds and randomization}
Already with one page, every deterministic completing algorithm has
competitive ratio at least two. A direct adversary releases each next
request shortly after the algorithm serves the previous one. The
one-page offline formula decides independently whether each arrival
gap should be crossed by one batch or split by an extra access. Comparing
these costs yields the lower bound, including against a fixed additive
constant; Section~\ref{sec:proof-lower} gives the full argument. The matching upper bound is equally simple: when there is only one
page, serve each pending episode exactly one unit after its first
arrival. A union of length-one intervals beginning at requests is a
samplewise offline lower certificate, while these first-arrival
intervals are disjoint across online episodes. Thus the best
deterministic factor for one page is exactly two
(Theorems~\ref{thm:deterministic-lower-two} and
\ref{thm:single-page-upper-two}). The five-factor multi-page bound
and~\eqref{eq:reverse-story} address different statements.

Randomized list update improves on deterministic reordering in the
classical problem \cite{albers1995}. Here, a randomized service or
reordering decision can change the future sequence of physical accesses.
An oblivious timed request sequence therefore does not automatically
produce an oblivious access sequence for a classical randomized list
algorithm. A transfer must control that dependence or analyze the
timing and reordering together.

One simple randomization does not improve the present certificate. Draw
a positive global scale $B$ independently of the input and run the
corresponding deterministic threshold. Averaging
Theorem~\ref{thm:main} gives
\begin{equation}\label{eq:random-scale-story}
 \E[\ALG_B]\le
 3(1+\E[B])C_O+2(1+\E[B^{-1}])D_O.
\end{equation}
Both coefficients could be below five only if
$\E[B]<2/3$ and $\E[B^{-1}]<3/2$, contradicting
$\E[B]\,\E[B^{-1}]\ge1$. The best uniform factor furnished by this
inequality is therefore five, attained by the deterministic scale
$2/3$. This is a limitation of the certificate, not a lower bound on
randomized algorithms or on a sharper analysis of random scales.

\subsection{What the comparison leaves to improve}
The main algorithm uses one clock per pending page and one familiar
reordering rule. Its analysis isolates two costs: creating marked-rank
credit when the comparator acts, and paying unmarked online intervals
from offline waiting. Improving only the numerical threshold balances
these two losses but does not reduce their combined factor below five.
A sharper comparison or a different reordering rule is needed for a
better guarantee through this approach.

Offline optimization presents a complementary question. The exact
state space separates input length from the number of pages, while the
static-order example shows why choosing one permanent order is
insufficient. A separate offline approximation could exploit future
arrivals while retaining adaptive list positions; an improved online
algorithm would also improve the present offline guarantee.

The technical part now establishes every comparison used above. It
first proves the event inequalities and threshold analysis, then gives
the exact offline construction and complexity reduction, extends the
interval charge, and verifies the lower bounds and failure examples.

%% file: sections/technical_online.tex
\section{The online comparison}
\label{sec:proof-online}

The proof separates list reordering from the timing rule. We first compare
the access costs of a move-to-front schedule and an arbitrary feasible
schedule. We then use the threshold rule to charge the remaining accesses
to the comparator's waiting cost. Throughout this section, $A$ denotes the
algorithm's list and $O$ the comparison list. Their initial permutations
are identical. We write $r_A(p)$ and $r_O(p)$ for the current ranks of page
$p$, and let $C_O$ include both accesses and paid adjacent exchanges.

\subsection{A coupling for delayed move-to-front schedules}

An episode of $A$ starts when a request arrives to a page with no pending
requests and ends when $A$ next accesses that page. Its oldest request is a
particular occurrence in the input, not merely a page identity. Mark a page
during its current episode once $O$ serves that oldest request. Retain the
mark until $A$ ends the episode. In particular, later arrivals to a marked
page do not remove its mark, even if they become pending in $O$.

For the analysis, reveal all arrivals at a timestamp first, then process
the actions of $O$, and finally those of $A$. Actions within each schedule
retain their original order. The two permutations are independent, so
this interleaving changes neither schedule's cost or behavior. It only
specifies the marks at simultaneous accesses. A new episode starts
unmarked because its oldest request has just arrived. Endpoints do not
affect any waiting-time integral used below.

Call an access of $A$ \emph{unmarked} if its page is unmarked immediately
before that access. Let $B_A$ be the sum of the access ranks at these
accesses. This quantity measures access cost, not the number of accesses.

\begin{lemma}[Move-to-front coupling]
\label{lem:tech-mtf-coupling}
Let $A$ be any completing schedule that clears a page's pending requests
at each access, moves the accessed page to the front, and makes no paid
exchanges. Its timing decisions may be arbitrary. For every feasible
comparison schedule $O$ with the same initial list,
\[
 C_A\le 3C_O+2B_A.
\]
The comparator may perform proactive paid adjacent exchanges and choose
any permitted free forward move after each access.
\end{lemma}

\begin{proof}
Let $I(A,O)$ count unordered pairs of pages whose relative orders differ
between the lists. Use the potential
\begin{equation}
 \Phi=I(A,O)+2\sum_{p\text{ marked}}r_O(p).
 \label{eq:tech-marked-potential}
\end{equation}
It is nonnegative and initially zero. Arrivals do not change it: an arrival
either joins an existing episode without changing its mark or starts an
unmarked episode. Passage of time also leaves it unchanged.

Consider an access by $A$ to a page $p$ of rank $k$, and write
$\ell=r_O(p)$. Let $a$ be the number of pages preceding $p$ in both lists.
Among the $k-1$ pages crossed by the move to front, exactly $a$ create an
inversion and the other $k-1-a$ remove one. Thus
\begin{equation}
 \Delta I=2a-(k-1),\qquad a\le\ell-1.
 \label{eq:tech-mtf-inversions}
\end{equation}
If $p$ is marked, ending its episode removes the term $2\ell$ from the
potential. The ranks and marks of other pages in $O$ do not change, so
\[
 k+\Delta\Phi
 =k+2a-(k-1)-2\ell\le-1\le0.
\]
If $p$ is unmarked, the marked-rank term is unchanged and
$\Delta I\le k-1$, giving
\[
 k+\Delta\Phi\le2k-1\le2k.
\]
Consequently all accesses of $A$ satisfy
\begin{equation}
 C_A+\sum_{\text{actions of }A}\Delta\Phi\le2B_A.
 \label{eq:tech-alg-event-sum}
\end{equation}

Next consider an access by $O$ to a page $p$ of rank $\ell$, including its
chosen free forward move. If this access first serves the oldest request
of $A$'s current episode for $p$, mark $p$ before accounting for its free
move. This increases the second term of~\eqref{eq:tech-marked-potential}
by $2\ell$. A forward move through $h\le\ell-1$ positions increases the
inversion count by at most $h$. When $p$ is marked, its rank decreases by
$h$, whereas at most $h$ other marked pages increase their ranks by one.
Hence the sum of marked ranks cannot increase during that move. The total
potential increase is at most $2\ell+h\le3\ell$.

If $p$ was already marked, the same argument applies without the new-mark
cost. If $p$ remains unmarked, its forward move increases the marked-rank
term by at most $2h$ and the inversion count by at most $h$. Its total
increase is again at most $3\ell$. This case also covers comparison
accesses that serve no request from a current episode of $A$.

Finally, a paid adjacent exchange by $O$ changes the inversion count by at
most one. Its change in the marked-rank term is zero when both exchanged
pages, or neither, are marked; otherwise its magnitude is two. Each paid
exchange therefore increases $\Phi$ by at most three. Summing comparison
actions gives
\begin{equation}
 \sum_{\text{actions of }O}\Delta\Phi\le3C_O.
 \label{eq:tech-opt-event-sum}
\end{equation}
Combining~\eqref{eq:tech-alg-event-sum}
and~\eqref{eq:tech-opt-event-sum}, and using the zero initial potential,
we obtain
\[
 C_A+\Phi_{\mathrm{final}}\le3C_O+2B_A.
\]
Dropping the nonnegative final potential proves the claim. No condition on
the final permutations is needed.
\end{proof}

\begin{proof}[Proof of Proposition~\ref{prop:temporal-transfer}]
By Lemma~\ref{lem:tech-mtf-coupling} and the two temporal bounds,
\[
 \ALG=C_A+D_A\le(1+\beta)C_A
 \le3(1+\beta)C_O+2(1+\beta)B_A
 \le3(1+\beta)C_O+\frac{2\kappa(1+\beta)}{\beta}D_O.
\]
For $\beta=2\kappa/3$, both coefficients equal $3+2\kappa$.
If $D_A=B_A=0$, Lemma~\ref{lem:tech-mtf-coupling} instead gives
$\ALG=C_A\le3C_O$ for the fixed comparator. Applying the respective
bound on every finite input against every feasible comparator proves
the competitive guarantees, since $A$ is feasible and completing.
The initial potential is zero and the
final potential nonnegative, so no additive term is needed.
\end{proof}

\subsection{Threshold equality and delay charging}

Fix $\beta>0$. The scaled rank-threshold algorithm accesses a pending page
$p$ when its oldest pending age reaches $\beta r_A(p)$ and moves it to the
front. After each access it recomputes all thresholds. If several pages
are eligible, it may select any one of them.

\begin{lemma}[Exact threshold and finite completion]
\label{lem:tech-timer-equality}
Every access of the scaled rank-threshold algorithm occurs at age exactly
$\beta r_A(p)$. Every episode beginning at $\alpha$ ends by
$\alpha+\beta m$, where $m$ is the number of pages. In particular,
\[
 D_A=\beta C_A,\qquad \ALG=(1+\beta)C_A.
\]
\end{lemma}

\begin{proof}
A new episode starts at age zero and has a positive threshold. Moving any
other page to the front leaves $p$'s rank unchanged or increases it by one;
it never decreases that rank. Between discrete events the episode's age
increases continuously. The algorithm acts at the first threshold
equality, and an intervening access can only increase another pending
page's threshold. Therefore no pending age exceeds its current threshold,
and every actual access has equality. This reasoning also covers ties:
serving one eligible page may make another temporarily ineligible, but
cannot create an overshoot.

Every rank is at most $m$, so an episode cannot persist beyond age
$\beta m$. Each access clears at least one request that is never cleared
again, and hence a finite input has only finitely many accesses. This also
rules out an infinite same-time action loop. Summing the threshold
equalities over all completed episodes gives the two cost identities.
\end{proof}

\begin{lemma}[Charging unmarked episodes]
\label{lem:tech-unmarked-delay}
For the scaled rank-threshold algorithm and every feasible comparator,
\[
 \beta B_A\le D_O.
\]
\end{lemma}

\begin{proof}
Consider an unmarked algorithm episode for page $p$, beginning at $\alpha$
and ending at $t$. Its oldest request has not been served by $O$ before the
algorithm's access. Thus $p$ is pending in $O$ throughout $[\alpha,t)$.
The interval's length is exactly $\beta$ times the algorithm's access
rank by Lemma~\ref{lem:tech-timer-equality}.

Distinct algorithm episodes for the same page occupy disjoint intervals,
apart from endpoints. Their unmarked intervals can therefore be charged
without repetition to that page's pending time in $O$. Since the
comparator's total delay is the integral of its number of pending pages,
summing over pages gives the claimed inequality. The argument permits
several algorithm episodes to lie inside one longer comparison episode.
\end{proof}

\begin{theorem}[Scaled rank-threshold guarantee]
\label{thm:tech-scaled-threshold}
For every $\beta>0$, every finite input, and every feasible comparison
schedule $O$,
\begin{equation}
 \ALG_\beta\le
 3(1+\beta)C_O+\frac{2(1+\beta)}{\beta}D_O.
 \label{eq:tech-scaled-bound}
\end{equation}
Thus its strict competitive factor is at most
$ (1+\beta)\max\{3,2/\beta\}$. Taking $\beta=2/3$ gives factor five;
taking $\beta=1$ gives factor six.
\end{theorem}

\begin{proof}
Lemmas~\ref{lem:tech-timer-equality} and
\ref{lem:tech-unmarked-delay} verify the temporal assumptions of
Proposition~\ref{prop:temporal-transfer} with $\kappa=1$.
That proposition gives~\eqref{eq:tech-scaled-bound} and its constants.
The empty input has zero cost and satisfies the same statement.
\end{proof}

For linear delay, the rule has a direct event implementation. Its next
possible service time is the minimum of $\alpha_p+\beta r_A(p)$ over
pending pages, unless an arrival occurs earlier. A scan of the list
updates ranks and finds this minimum in $O(m)$ arithmetic operations per
event. There are $N$ request arrivals and at most $N$ accesses, so
chronologically presented input can be processed in $O(Nm)$ arithmetic
operations and $O(m)$ working memory. For $\beta=2/3$ and rational arrival
times, all service times and comparisons have polynomial bit complexity.
The implementation uses its own timers and does not need to know $N$ or
the last arrival time.

%% file: sections/technical_offline.tex
\section{Offline computation and complexity}
\label{sec:proof-offline}

This section concerns the linear-delay model. Let $N$ be the number of
requests and let $m\ge1$ be the number of pages. Requests are supplied in
nondecreasing arrival-time order. Repeated requests at the same time are
allowed, and all such requests are revealed before any action at that time.
The terminal permutation is unrestricted.

\subsection{Arrival epochs and a finite-state optimum}

\begin{lemma}[Arrival-layer normal form]
\label{lem:offline-arrival-normal-form}
Every finite feasible schedule can be replaced, without increasing its
cost, by one whose accesses and paid exchanges all occur at request-arrival
epochs. At each epoch there are at most $m$ useful accesses. The replacement
retains the freedom to choose any legal forward position after an access.
\end{lemma}
\begin{proof}
An access to a page with no pending request is unnecessary. If its rank is
$r$ and its subsequent free move advances it by $h$ positions, replace the
access and free move by $h$ paid adjacent exchanges. This produces the same
permutation and pending set and costs $h\le r-1<r$. An access followed by no
move can simply be deleted. Thus we may assume that every access clears a
nonempty pending set.

Consider consecutive distinct arrival epochs $t_j<t_{j+1}$. Move all
operations in the open interval $(t_j,t_{j+1})$ to immediately after the
operations originally performed at $t_j$, retaining their relative order.
There is no intervening arrival, so every access clears the same requests,
all intermediate permutations remain the same, and every free or paid move
remains legal. Access and exchange costs are unchanged, while every affected
request is served weakly earlier. Hence waiting costs do not increase.
Paid exchanges before the first arrival may likewise be postponed to that
arrival, before the schedule's original actions. All operations after the
last arrival may be moved to that arrival. Applying these transformations
gives the claimed normal form.

After a page is accessed at one epoch, it has no pending request and no new
request is revealed until the next epoch. Thus there is at most one useful
access to each page at an epoch. No part of the argument requires a free
move to reach the front: the original forward destination is preserved.
\end{proof}

For a schedule, let $P(t)$ be its pending-page set. Its linear waiting cost
is
\begin{equation}
 D=\int |P(t)|\,dt.
 \label{eq:offline-occupancy-cost}
\end{equation}
Indeed, each pending episode contributes one to the integrand throughout
its duration and contributes exactly that duration to the objective.
Consequently, if waiting is charged as time passes, the remaining cost from
a given time depends only on the permutation and the pending-page set.
The ages of currently pending episodes need not be stored: their past
waiting has already been paid.

\begin{theorem}[Exact parameterized algorithm]
\label{thm:offline-fpt}
Set $S=m!2^m$. The offline optimum can be computed in
$O(NSm\log S)$ arithmetic operations, using $O(Sm)$ space.
The algorithm permits arbitrary paid adjacent exchanges and all legal
free forward moves. For rational arrival times, it is a fixed-parameter
tractable exact algorithm with parameter $m$, with polynomial additional
dependence on the binary input length.
\end{theorem}
\begin{proof}
First observe that optional free forward moves need not appear explicitly
in an optimum. An access at rank $r$ followed by a free forward move of
$h$ positions can instead be implemented by $h$ paid forward exchanges
\emph{before} the access and then an access at rank $r-h$. The total cost
is still $r$, the access occurs at the same time, and the final permutation
and pending set are identical. All other operations can be retained in
their original order. Thus the optimum is unchanged if every forward
movement is paid and accesses themselves leave the permutation unchanged.

Construct a directed graph with one vertex $(\pi,P)$ for each permutation
$\pi$ of the $m$ pages and each subset $P$ of pages. Its arcs encode
instantaneous actions:
\begin{itemize}
\item An adjacent exchange in $\pi$ has cost one and leaves $P$ unchanged.
\item Accessing $p\in P$ has cost $\operatorname{rank}_{\pi}(p)$, removes
      $p$ from $P$, and leaves the permutation unchanged.
\end{itemize}
Every vertex has $O(m)$ outgoing arcs. A path represents a legal sequence
of actions at one time; conversely, the preceding transformation represents
every useful instantaneous action sequence by a path of the same cost,
including sequences originally using optional free forward moves.
All arc costs are positive. One could include all free-forward access arcs
explicitly, but they are redundant for shortest-path distances.

Let $t_1<\cdots<t_K$ be the distinct arrival epochs, with $K\le N$, and
let $R_j$ be the set of pages requested at $t_j$. Let $F_j(\pi,P)$ be the
minimum cost just after the actions at $t_j$, with waiting already charged
through $t_j$. To pass from epoch $j-1$ to epoch $j$, each state $(\pi,P)$
first pays $(t_j-t_{j-1})|P|$ and then activates the new requests. Thus its
cost is a candidate seed for state $(\pi,P\cup R_j)$. Take the minimum
when several predecessor states give the same seed. Apply multi-source
Dijkstra to these seeds to obtain $F_j$ at every vertex. For the first
epoch the unique predecessor is the given initial permutation with empty
pending set and zero cost, and the first waiting increment is zero.

Lemma~\ref{lem:offline-arrival-normal-form} and
\eqref{eq:offline-occupancy-cost} show inductively that this recurrence
represents every normalized feasible schedule and assigns it its exact
cost. The answer is
\[
 \min_{\pi}F_K(\pi,\varnothing).
\]
In particular, all requests are completed; no constraint is imposed on the
terminal list. Empty input has value zero.

The graph has $S$ vertices and $O(Sm)$ arcs. Multi-source Dijkstra with a
binary heap costs $O(Sm\log S)$ per epoch, and two cost vectors suffice
between epochs. The graph and the two vectors use $O(Sm)$ space.
The displayed time bound counts arithmetic operations. For rational
timestamps encoded in binary, all relevant costs are sums of integer
access costs and integer multiples of differences of input timestamps.
They have polynomial encoding length in the input size; exact additions
and comparisons therefore add only a polynomial bit-complexity factor.
The exponential factor depends only on $m$.
\end{proof}

The state compression is specific to linear waiting, or to other delay
rules for which the accrued cost has an appropriate memoryless state
description. It is not asserted for arbitrary age-dependent episode
penalties. In particular, nonlinear penalties in the online extension do
not automatically admit the same offline state space.

\subsection{NP-completeness for an unrestricted number of pages}

\begin{theorem}[Offline decision complexity]
\label{thm:offline-npcomplete}
Given rational arrival times and a rational bound $B$, deciding whether
there is a complete schedule of cost at most $B$ is NP-complete.
\end{theorem}
\begin{proof}
For membership in NP, use the arrival-layer normal form. There are at most
$m$ useful accesses at each epoch. Between consecutive accesses, replace
the paid exchanges by a shortest sequence of adjacent swaps producing the
same endpoint permutation. Such a sequence has length at most
$m(m-1)/2$. The same replacement applies before the first and after the
last access at the epoch. The resulting certificate has $O(Nm^3)$
elementary actions. The free forward destination of each access is recorded
explicitly. Legality, completion, and the total rational cost can all be
checked in polynomial time.

For NP-hardness, reduce from classical offline list update, which is
NP-hard~\cite{ambuhl2000}. Let its initial list contain $m$ pages and its
request sequence have length $N$. Put request $i$ at time $iH$, where
$H=Nm+1$. Serving every request immediately without changing the list is
feasible and costs at most $Nm$. If a request were still pending at the
next arrival, its page episode would eventually pay waiting cost at least
$H>Nm$. Such a schedule cannot be optimal. Consequently, an optimum serves
each request before the next one arrives.

Apply Lemma~\ref{lem:offline-arrival-normal-form}. Each request is now
served at its own arrival, with zero waiting, and the same ordering and
list operations form a classical list-update schedule. Conversely, any
classical schedule can be executed at these arrival epochs with exactly
its original cost and zero waiting. The two optimal values are equal, and
the construction uses timestamps of polynomial encoding length.

The classical cost convention in this reduction agrees with ours. In the
full-cost model an access at rank $r$ followed by a free forward move of
$h$ positions can be replaced by $h$ paid swaps before the access, followed
by an access at rank $r-h$, with the same total cost and final permutation.
Thus allowing optional free forward moves does not invalidate the classical
offline hardness statement. Also, changing classical access cost from
$r-1$ to $r$ adds exactly $N$ to every schedule, so the partial- and
full-cost offline hardness statements transfer directly.
\end{proof}

\subsection{Exact fixed-list computation and its limitation}

\begin{proposition}[Fixed permutation]
\label{prop:offline-fixed-order}
Suppose no list changes are permitted. For a page of fixed rank $c$ with
distinct arrival times $t_1<\cdots<t_q$, its optimum cost is zero if $q=0$
and otherwise equals
\begin{equation}
 G_p(c)=c+\sum_{i=1}^{q-1}\min\{t_{i+1}-t_i,c\}.
 \label{eq:offline-fixed-order}
\end{equation}
Summing these quantities over pages gives the fixed-list optimum, which
can be computed in $O(N+m)$ time from chronologically ordered input.
\end{proposition}
\begin{proof}
Pages are independent when ranks are fixed. An access clears all pending
requests of its page, so its batches consist of consecutive arrivals of
that page. Move each access to the last arrival of its batch, reducing
waiting without changing the partition. The cost of a batch is $c$ plus
the sum of the gaps internal to that batch. Starting with one access cost
$c$, each consecutive gap independently either remains inside a batch,
at the cost of its length, or is cut, at the cost of one additional access
$c$. Choosing the cheaper option proves the formula. Equal-time requests
can be collapsed, since they are always cleared together.
\end{proof}

\begin{proposition}[Best freely chosen static permutation]
\label{prop:offline-best-free-static}
If an initial permutation may be chosen at no cost and must then remain
fixed, an optimal choice and schedule can be computed in polynomial time.
\end{proposition}
\begin{proof}
Compute $G_p(r)$ from \eqref{eq:offline-fixed-order} for each page $p$ and
each possible rank $r\in\{1,\ldots,m\}$. Assigning pages to distinct ranks
is a minimum-cost perfect matching in the bipartite graph of pages and
ranks, where edge $(p,r)$ has cost $G_p(r)$. The resulting permutation,
with the per-page partitions in Proposition~\ref{prop:offline-fixed-order},
is optimal. This statement does not include the cost of changing a given
initial permutation: adding that permutation's inversion cost would not
be the separable assignment objective used here.
\end{proof}

\begin{proposition}[Static permutations cannot give a constant factor]
\label{prop:offline-static-gap}
Even if its initial permutation is chosen for free, the best fixed-list
schedule can be a factor approaching $(m+1)/2$ more expensive than the
unrestricted offline optimum.
\end{proposition}
\begin{proof}
Start the unrestricted schedule in order $(p_1,\ldots,p_m)$. For an integer
$K\ge1$, give $K$ consecutive requests to $p_1$, then $K$ to $p_2$, and so
on through $p_m$. Space all arrivals by $H>m$, so every consecutive gap of
a given page exceeds every possible rank. Each page has $K$ requests, and
Proposition~\ref{prop:offline-fixed-order} gives cost $Kr$ if it is assigned
rank $r$. Thus every freely chosen static permutation costs
$Km(m+1)/2$.

A feasible dynamic schedule serves every request immediately and moves
the accessed page to the front. The first access in phase $j$ has rank $j$;
the other $K-1$ accesses of that phase have rank one. Its total cost is
$mK+m(m-1)/2$. Hence the static optimum divided by unrestricted $\OPT$ is
at least
\[
 \frac{Km(m+1)/2}{mK+m(m-1)/2}
 =\frac{K(m+1)}{2K+m-1}.
\]
Letting $K$ grow proves the claim. The construction has zero waiting in
the displayed dynamic comparison schedule.
\end{proof}

%% file: sections/technical_extensions.tex
\section{Extensions of the delay charge}
\label{sec:proof-extensions}

The list comparison in Lemma~\ref{lem:tech-mtf-coupling} does not involve
waiting costs. Proposition~\ref{prop:temporal-transfer} requires the
temporal bounds $D_A\le\beta C_A$ and $\beta B_A\le\kappa D_O$,
together with feasibility and completion. Without deadlines, our rules
satisfy exact threshold equality at every service. With deadlines,
services may occur earlier; only unmarked services need hit their
thresholds exactly for the interval charge. We verify these properties
for each model below. Every access still clears one page, and both
schedules face the same input penalties and deadlines.

\subsection{Superadditive penalties for elapsed waiting time}

For each page $p$, fix a function
$f_p:[0,\infty)\to[0,\infty)$ which is continuous, nondecreasing, and
unbounded, with $f_p(0)=0$. Assume also that
\begin{equation}
 f_p(x+y)\ge f_p(x)+f_p(y)
 \quad\text{for all }x,y\ge0.
 \label{eq:tech-superadditivity}
\end{equation}
An episode beginning at $\alpha$ and served at $t$ incurs penalty
$f_p(t-\alpha)$. Since the function is nondecreasing and is common to all
requests on page $p$, this is the maximum of the requests' individual
penalties in that episode. Functions may differ between pages. The
algorithm triggers when $f_p(t-\alpha_p)$ first reaches
$\beta r_A(p)$.

\begin{lemma}[Superadditive interval charge]
\label{lem:tech-superadditive-charge}
Under these assumptions, every algorithm access occurs with penalty
$\beta r_A(p)$, every finite input is completed, and
\[
 \beta B_A\le D_O.
\]
Here $D_O$ is the sum of the comparator's episode penalties.
\end{lemma}

\begin{proof}
Continuity gives equality at the first threshold crossing. As in
Lemma~\ref{lem:tech-timer-equality}, other move-to-front operations only
increase a pending page's rank, so they cannot cause a penalty to overshoot
its threshold. This also applies when $f_p$ has flat portions. Since
$f_p$ is unbounded, an episode ends no later than the first age at which
$f_p$ reaches $\beta m$. The number of accesses remains at most the
number of requests, so completion is finite.

Every unmarked algorithm episode lies within a single comparison episode
of the same page: its oldest request is pending throughout its interval
in the comparator. Group the unmarked intervals by their containing
comparison episode. If their lengths are $x_1,\ldots,x_s$ and that
comparison episode has length $L$, the intervals are disjoint and
$\sum_i x_i\le L$. Superadditivity and monotonicity imply
\[
 \sum_{i=1}^s f_p(x_i)
 \le f_p\!\left(\sum_{i=1}^s x_i\right)
 \le f_p(L).
\]
Each term on the left is $\beta$ times its algorithm access rank.
Summing these inequalities over comparison episodes proves
$\beta B_A\le D_O$.
\end{proof}

\begin{theorem}[Superadditive episode penalties]
\label{thm:tech-superadditive}
For the functions above, scaled rank-threshold MTF satisfies
\[
 \ALG_\beta\le3(1+\beta)C_O+
 \frac{2(1+\beta)}{\beta}D_O
\]
against every feasible comparator. In particular, $\beta=2/3$ gives a
strict competitive factor of five.
\end{theorem}

\begin{proof}
The marked-rank coupling gives $C_A\le3C_O+2B_A$. By
Lemma~\ref{lem:tech-superadditive-charge},
$B_A\le D_O/\beta$, and exact threshold equality gives
$D_A=\beta C_A$. Combining these facts proves the result.
\end{proof}

Continuous nondecreasing unbounded convex functions with value zero at
zero satisfy~\eqref{eq:tech-superadditivity}. Indeed, for $x+y>0$,
convexity gives $f_p(x)\le x f_p(x+y)/(x+y)$ and the corresponding
inequality for $y$; adding proves superadditivity. Positive page-specific
linear weights are a special case. The theorem does not require strict
increase or strict convexity.

The interval inequality is substantive: Proposition~\ref{prop:concave-threshold-failure}
shows that the fixed-scale rule has unbounded ratio on concave powers,
already for one page. That statement concerns this rule, not every
possible algorithm for those penalties.

\subsection{Exogenous rates on the time line}

For each page $p$, let $w_p:[0,\infty)\to[0,\infty)$ be an exogenous
locally integrable rate, shared by the two schedules. An episode
$[\alpha,t]$ has penalty
\begin{equation}
 F_p(\alpha,t)=\int_\alpha^t w_p(s)\,ds.
 \label{eq:tech-rate-penalty}
\end{equation}
Nonnegativity makes this the maximum of the integrals from the individual
request arrival times to $t$, since the oldest arrival gives the largest
integration interval.
Rates may vary over time and need not be known in advance. The algorithm
observes the penalty accrued by each pending page and acts when it first
reaches $\beta r_A(p)$. To ensure completion of every finite input, assume
\begin{equation}
 \int_u^\infty w_p(s)\,ds=\infty
 \quad\text{for every page }p\text{ and every finite }u\ge0.
 \label{eq:tech-rate-tail}
\end{equation}
Local integrability makes every accrued-penalty function continuous.
Condition~\eqref{eq:tech-rate-tail} permits intervals of zero rate but
rules out a pending page whose remaining lifetime penalty is bounded.

\begin{lemma}[Charging common calendar rates]
\label{lem:tech-calendar-charge}
For the rate model, the algorithm completes every finite input and
satisfies $D_A=\beta C_A$ and $\beta B_A\le D_O$.
\end{lemma}

\begin{proof}
Continuity and nondecreasing accrued penalties give exact threshold
equality, because other move-to-front operations cannot lower a pending
page's threshold. Condition~\eqref{eq:tech-rate-tail} ensures that an
episode starting at $\alpha$ reaches penalty $\beta m$ in finite time,
so it must be served by then. Summing equality at service gives
$D_A=\beta C_A$.

As before, each unmarked algorithm episode occupies an interval during
which the same page is pending in $O$. These intervals are disjoint for
each page. Since the rate is nonnegative and common to the two schedules,
their penalties sum to at most that page's total pending-rate integral
in $O$. Equivalently,
\[
 \begin{split}
 \beta B_A
 &=\sum_p\int_{\substack{s\text{ in an unmarked}\\
                         \text{algorithm episode of }p}}
                 w_p(s)\,ds\\
 &\le\sum_p\int_0^\infty
       w_p(s)\,\mathbf{1}\{p\text{ is pending in }O\text{ at }s\}\,ds
 =D_O.
 \end{split}
\]
The equality on the right follows by summing the integrals of the
comparator's disjoint episodes for each page.
\end{proof}

\begin{theorem}[Common exogenous page rates]
\label{thm:tech-calendar-rates}
Under~\eqref{eq:tech-rate-penalty}--\eqref{eq:tech-rate-tail}, the scaled
rank-threshold algorithm satisfies
\[
 \ALG_\beta\le3(1+\beta)C_O+
 \frac{2(1+\beta)}{\beta}D_O.
\]
In particular, $\beta=2/3$ gives a strict factor of five without knowledge
of future rates.
\end{theorem}

\begin{proof}
Apply Lemma~\ref{lem:tech-mtf-coupling} and
Lemma~\ref{lem:tech-calendar-charge}, then use
$\ALG_\beta=(1+\beta)C_A$.
\end{proof}

The rule uses only presently accrued penalties. Its continuous-time
description assumes access to their first hitting times; the number of
accesses is still at most $N$. For arbitrary functions or rates, this is
not a claim of polynomial bit complexity for evaluating those hitting
times. The rates must be external to the two schedules: replacing a
common calendar rate by a rate reset or changed separately by each
algorithm's actions would invalidate the interval comparison. Likewise,
the elapsed-time theorem uses one fixed function per page, rather than
arbitrary functions for different occurrences of that page.

\subsection{Hard deadlines and bounded request heterogeneity}
\label{sec:tech-deadline-heterogeneous}

We next allow both hard deadlines and some heterogeneity among the
penalties of requests for the same page. The result is not a consequence
of applying the previous page-specific theorem independently to each
request: a single page access must clear all of them, and the maximum
penalty need not belong to the oldest request.

For each request $v$, let $p(v)$ be its page, $a_v$ its arrival,
$d_v\in[a_v,\infty]$ its deadline, and
$f_v:[0,\infty)\to[0,\infty)$ its waiting penalty. A page access
at time $t$ clears all requests then pending at that page and incurs the
penalty $\max_v f_v(t-a_v)$ over those requests. Every request must be
served no later than its finite deadline. Infinite deadlines simply mean
that the request has no hard deadline. An online algorithm learns
$d_v$ and $f_v$ when request $v$ arrives.

Fix a known $\kappa\ge1$. For each page $p$, assume there is a continuous,
nondecreasing, unbounded, superadditive function $F_p$, with $F_p(0)=0$,
such that every request $v$ for $p$ satisfies
\begin{equation}\label{eq:envelope}
 F_p(x)\le f_v(x)\le\kappa F_p(x)
 \qquad (x\ge0).
\end{equation}
All $f_v$ are assumed continuous, nondecreasing, and zero at zero.
The functions $f_v$ need not be identical within a page.

For a pending page $p$ write
$H_p(t)=\max_{v\in P_A(p,t)}f_v(t-a_v)$ and let $\delta_p(t)$
be the earliest deadline among its pending requests. Fix $\beta>0$.
Serve $p$ at the first time that either
$H_p(t)=\beta r_A(p)$ or $t=\delta_p(t)$, then move $p$ to the front.
Recompute eligibility after each access. At a common timestamp, reveal
arrivals first and repeatedly clear every page that is eligible by
either condition; all pending deadlines must be met.

\paragraph{Why no new marks are necessary.}
For one-page clearing, the two natural marking conventions are
equivalent. If the comparator has served any request belonging to an
online episode, it has also served that episode's oldest request:
it either cleared the oldest request in the same access or had already
cleared it earlier. Conversely, serving the oldest request itself
creates a mark. Thus the mark definition and the coupling of
Lemma~\ref{lem:tech-mtf-coupling} apply unchanged, including at
simultaneous deadlines when comparator actions are accounted for first.

\begin{theorem}[Maximum-delay list update with deadlines and heterogeneity]
\label{thm:hybrid-extension}
Under~\eqref{eq:envelope}, the earliest-deadline-or-threshold MTF
algorithm with $\beta=2\kappa/3$ is deterministic and strictly
$(3+2\kappa)$-competitive against the unrestricted clairvoyant
comparator, including the cost of completing every finite input.
Its competitive ratio is independent of the number of pages and
requests. In particular, for one fixed superadditive penalty function
per page ($\kappa=1$), arbitrary per-request hard deadlines can be added
without worsening the strict factor five.
\end{theorem}
\begin{proof}
During an online episode the oldest arrival remains unchanged. The
functions defining $H_p$ are continuous between arrivals; at an arrival
a new request contributes $f_v(0)=0$, hence $H_p$ does not jump up.
An access of another page via MTF can only increase or preserve the
rank of a still-pending page. Thus the current penalty never exceeds
$\beta r_A(p)$ before its service. In particular,
\begin{equation}\label{eq:hybrid-online-bound}
 D_A\le\beta C_A.
\end{equation}
All requests are completed: by~\eqref{eq:envelope}, an episode's
oldest request eventually reaches penalty at least $F_p(t-a)$, which
is unbounded, whereas its threshold is at most $\beta m$.
Deadlines can only trigger earlier accesses. Since each access clears
at least one request, only finitely many accesses occur on finite
inputs, including at tied timestamps.

Use the marks in Lemma~\ref{lem:tech-mtf-coupling}. If $A$ accesses
page $p$ because a deadline $d_v$ is reached, the comparator must
already have served request $v$; order comparator actions before online
actions at the same timestamp for bookkeeping. Thus $p$ is marked.
Consequently, every unmarked online access was caused by exact threshold
hitting and incurs penalty $\beta r_A(p)$.

Fix an unmarked online episode $I=[a,t]$ of page $p$, of length $x$.
None of its requests was served by the comparator before its end, so
its oldest request remains pending in the comparator throughout $I$.
The interval $I$ therefore lies within one comparator episode $J$ of
page $p$. The penalty at this online service is at most $\kappa F_p(x)$:
every request it serves arrived at or after $a$, has age at most $x$,
and obeys~\eqref{eq:envelope}. Hence
$\beta r_A(p)\le\kappa F_p(x)$ for every unmarked service.

Now fix a comparator episode $J$ of page $p$ having length $L$.
The unmarked online episodes $I_1,\ldots,I_s$ contained in $J$
are pairwise disjoint. Therefore $\sum_j|I_j|\le L$, and
superadditivity and monotonicity give
\[
 \beta\sum_{j=1}^s r_A(p;I_j)
 \le \kappa\sum_{j=1}^s F_p(|I_j|)
 \le \kappa F_p\left(\sum_{j=1}^s|I_j|\right)
 \le \kappa F_p(L).
\]
At the comparator access ending $J$, its oldest request has age $L$;
therefore the comparator's maximum penalty for this episode is at least
$F_p(L)$ by~\eqref{eq:envelope}. Summing across comparator episodes
and pages yields
\begin{equation}\label{eq:hybrid-charge}
 \beta B_A\le\kappa D_O.
\end{equation}

Equations~\eqref{eq:hybrid-online-bound}
and~\eqref{eq:hybrid-charge} verify the assumptions of
Proposition~\ref{prop:temporal-transfer}. Its choice
$\beta=2\kappa/3$ proves the strict factor $3+2\kappa$.
\end{proof}

\begin{corollary}[Hard deadlines without waiting penalties]
\label{cor:deadline-only-3}
In the same single-page-access list-update model, assume every request
has a finite hard deadline, and there is no waiting penalty before the
deadline. Access a page at the earliest pending deadline for that page,
clear all its requests, and move it to the front. This algorithm is
strictly $3$-competitive against an unrestricted feasible offline
schedule.
\end{corollary}
\begin{proof}
Use the marks in Lemma~\ref{lem:tech-mtf-coupling}. Every online access
is triggered by some request's hard deadline. The comparator must have
served that request no later than the same timestamp, so every online
access is marked. Therefore $B_A=0$ and
$\ALG=C_A\le3C_O=3\OPT$.
\end{proof}

\paragraph{Relation to the request-specific counterexample.}
Proposition~\ref{prop:request-weights-failure} allows the ratio of
request-specific weights on one page to grow without bound. The theorem
above instead quantifies what is retained when such heterogeneity is
uniformly bounded: if $f_v(x)=w_vF_p(x)$ and
$w_v\in[w_p,\kappa w_p]$, take the baseline to be $w_pF_p$.
The algorithm needs a known bound $\kappa$ and the currently accrued
request penalties; no future arrivals are required. As with the
previous arbitrary-function guarantees, no generic polynomial bit-time
claim is made for evaluating arbitrary penalty functions.

\subsection{One-sided arrival-order domination without a common baseline}
\label{sec:tech-arrival-ordered}
The bounded-envelope theorem assumes a common superadditive function
for a page. There is a different structural condition that needs no
common baseline and tolerates arbitrarily large \emph{decreases} in
request severity over time. It only limits how much a later request can
exceed an earlier one.

\begin{theorem}[One-sided arrival-order domination]
\label{thm:arrival-ordered}
For each request $v$ let $f_v:[0,\infty)\to[0,\infty)$ be continuous,
nondecreasing, unbounded, superadditive, and satisfy $f_v(0)=0$.
Fix a known $\kappa\ge1$ and suppose that, for every two requests
$u,v$ to the same page with $a_u\le a_v$,
\begin{equation}\label{eq:arrival-ordered}
 f_v(x)\le\kappa f_u(x)\qquad\text{for every }x\ge0.
\end{equation}
(The condition applies in both directions at equal timestamps.)
Each request may also have an arbitrary hard deadline in
$[a_v,\infty]$. Accessing a page costs its rank plus the maximum of
the pending requests' current penalties. Access a page at the first
of a hard-deadline expiration or a hit of the threshold
$\beta r_A(p)$ by its maximum pending penalty, where
$\beta=2\kappa/3$, and move that page to the front.
This deterministic rule is strictly $(3+2\kappa)$-competitive against
the unrestricted clairvoyant comparator on every finite input.
In particular, if later requests never have larger penalty functions
than earlier requests ($\kappa=1$), the factor is $5$, with \emph{no}
bound on the ratio of their magnitudes.
\end{theorem}
\begin{proof}
For a pending page $p$, the maximum of its requests' accrued penalties
is continuous between arrivals; newly arriving requests have zero
penalty, so it cannot jump up at an arrival. An MTF access of another
page cannot lower $p$'s rank. Thus an access never pays more than
$\beta r_A(p)$ in waiting penalty, and every access not forced by a
deadline hits the threshold exactly. Every finite input is completed:
the penalty of any unserved request eventually exceeds $\beta m$.

Mark an online page episode at the first comparator access serving
any request in the episode, as in Lemma~\ref{lem:tech-mtf-coupling}.
Deadline-forced online accesses are marked because the comparator
must meet the same deadlines; consequently, every unmarked online
access occurs at exact threshold. Hence
\begin{equation}\label{eq:arrival-spatial}
 C_A\le3C_O+2B_A,\qquad D_A\le\beta C_A.
\end{equation}

Fix an unmarked online episode $I=[a,t]$ of page $p$.
Its oldest request remains pending in the comparator throughout $I$,
so $I$ lies in a single comparator episode $J=[b,s]$ for the same
page, with $b\le a$ and length $L=s-b$. Let $u$ be any request
arriving at the start $b$ of $J$. For every request $v$ cleared by
the online access ending $I$ we have $a_v\ge a\ge b$. Thus the
one-sided comparison~\eqref{eq:arrival-ordered}, followed by
monotonicity, gives
\[
  f_v(t-a_v)\le\kappa f_u(t-a_v)
   \le\kappa f_u(t-a)=\kappa f_u(|I|).
\]
Since $I$ ends unmarked, it is a threshold access, so
$\beta r_A(p;I)\le\kappa f_u(|I|)$.

For all unmarked online episodes $I_1,\ldots,I_h$ lying in $J$,
their interiors are disjoint and $\sum_j|I_j|\le L$. The same
comparator request $u$ and its superadditive function $f_u$ can be
used for every one of them. Therefore
\begin{align*}
 \beta\sum_{j=1}^h r_A(p;I_j)
 &\le\kappa\sum_{j=1}^h f_u(|I_j|)
 \le\kappa f_u\!\left(\sum_{j=1}^h |I_j|\right)
 \le\kappa f_u(L)
 \le\kappa D_O(J).
\end{align*}
Here the last inequality follows because $u$ is present at the
comparator access ending $J$, with age $L$; its individual penalty
is no more than the maximum penalty charged by that access.
Summing over pages and comparator episodes yields
$\beta B_A\le\kappa D_O$. Together with~\eqref{eq:arrival-spatial}, this verifies both
conditions of Proposition~\ref{prop:temporal-transfer}; choosing
$\beta=2\kappa/3$ gives the strict factor $3+2\kappa$.
\end{proof}

\begin{corollary}[Nonincreasing request weights]
\label{cor:nonincreasing-weights}
Fix, for each page $p$, a continuous, nondecreasing, unbounded,
superadditive function $F_p$ with $F_p(0)=0$. Let each request $v$
for page $p$ have penalty $w_vF_p(x)$ with $w_v>0$, where the
weights are nonincreasing in arrival order and are equal among
simultaneous requests of the same page. No bound on the ratio of
weights is needed. The deadline-or-threshold MTF algorithm of
Theorem~\ref{thm:arrival-ordered} is strictly $5$-competitive.
\end{corollary}
\begin{proof}
Each $f_v=w_vF_p$ satisfies the theorem's pointwise ordering
and superadditivity hypotheses.
\end{proof}

\paragraph{Why this does not contradict the negative example.}
Proposition~\ref{prop:request-weights-failure} uses request weights
that \emph{increase} rapidly with each new arrival. The theorem
covers the opposite direction, including arbitrarily steep decreases.
It does not prove a constant competitive ratio for arbitrary
request-specific maximum penalties.

%% file: sections/technical_lower.tex
\section{Lower bounds and limits of the threshold analysis}
\label{sec:proof-lower}

The statements in this section distinguish three different conclusions:
a lower bound for all deterministic algorithms, a lower bound for the
particular scaled threshold rule, and failures of that rule outside its
proved delay-function assumptions. They are not interchangeable.

\subsection{A deterministic problem lower bound}

\begin{theorem}[One-page deterministic lower bound]
\label{thm:deterministic-lower-two}
No deterministic completing algorithm has competitive ratio strictly below
two, even for one page with unit access cost and linear episode delay.
The statement remains true when a fixed additive constant is allowed.
\end{theorem}
\begin{proof}
Fix $1\le c<2$ and any finite additive constant $b$. Choose
$\varepsilon>0$ such that
$\delta=2-c-c\varepsilon>0$. We construct requests one at a time. The
first request arrives at $a_1=0$. Given the current finite prefix, consider
the algorithm's continuation if no further request arrives. Completion
requires it to access the pending page at a finite time $s_i\ge a_i$.
Choose the next arrival $a_{i+1}$ strictly after $s_i$ and at most
$s_i+\varepsilon$. The next arrival can be chosen rational. Determinism
ensures that the algorithm actually behaves as simulated through $s_i$.
After $N$ requests, stop the construction and let the algorithm complete
the last request normally. There is no end-of-input notification.

Let $\tau_i=s_i-a_i$. Each request is alone in its algorithm episode, so
the algorithm pays at least $\sum_{i=1}^N(1+\tau_i)$. Extra accesses, if
any, only increase its cost. The one-page optimum formula in
Proposition~\ref{prop:offline-fixed-order} gives
\[
 \OPT=1+\sum_{i=1}^{N-1}\min\{a_{i+1}-a_i,1\}
 \le 1+\sum_{i=1}^{N-1}\min\{\tau_i+\varepsilon,1\}.
\]
For every $x\ge0$,
\[
 1+x-c\min\{x+\varepsilon,1\}
 \ge 1+x-c\min\{x,1\}-c\varepsilon
 \ge 2-c-c\varepsilon=\delta.
\]
The last inequality follows separately on $x\le1$ and $x\ge1$.
Consequently
\[
 \ALG-c\OPT\ge (N-1)\delta+1-c.
\]
Taking $N$ sufficiently large makes this exceed $b$. Although the
construction is adaptive, for a fixed deterministic algorithm it specifies
a finite input sequence, as required for a deterministic lower bound.
\end{proof}

\begin{theorem}[Tight deterministic guarantee for one page]
\label{thm:single-page-upper-two}
For one page of unit access cost and linear per-episode maximum delay,
the algorithm that serves a pending episode exactly one time unit after
its first arrival is strictly $2$-competitive against the unrestricted
clairvoyant optimum. Together with
Theorem~\ref{thm:deterministic-lower-two}, the optimal deterministic
competitive factor for this single-page problem is exactly $2$.
\end{theorem}
\begin{proof}
For each request at time $a$, draw the interval $[a,a+1]$ and let
$U$ be their union. Consider any offline batch served at time $t$,
whose earliest request arrived at $a_{\min}$. All the intervals
of requests in the batch are contained in $[a_{\min},t+1]$.
The latter interval has length $1+t-a_{\min}$, exactly the offline
batch's access-plus-maximum-delay cost. Taking the union across all
batches and using subadditivity of length gives $|U|\le\OPT$.

Let $a_1,\ldots,a_K$ be the first arrival times of the online
algorithm's successive episodes. Each episode is cleared at
$a_j+1$. The next episode begins only after that clearing,
so the intervals $[a_j,a_j+1]$ have disjoint interiors and are all
contained in $U$. Therefore $|U|\ge K$. Each online episode costs
one access plus one unit of delay, and hence
$\ALG=2K\le2|U|\le2\OPT$. This applies to every finite input and
has no additive term. The deadline-free policy needs no signal
that the input has ended.
\end{proof}

\subsection{A lower bound for the scaled rank threshold}

\begin{proposition}[Reverse blocks]
\label{prop:scaled-threshold-lower}
For every fixed $\beta>0$ and every $m\ge1$, the scaled rank-threshold
algorithm has input sequences on which its cost divided by $\OPT$ is at
least
\[
 \frac{2(1+\beta)m}{m+1}.
\]
The construction can be repeated arbitrarily many times. Thus the rule's
competitive ratio uniform in $m$ is at least $2(1+\beta)$, including when
an instance-independent additive constant for each fixed $m$ is allowed.
In particular, the lower bound approaches $10/3$ at $\beta=2/3$ and four
at $\beta=1$.
\end{proposition}
\begin{proof}
Start from $(1,\ldots,m)$ and repeatedly request the block
$m,m-1,\ldots,1$. Separate consecutive arrivals by a fixed
$H>\beta m$. Each request is served before the next arrival. At every
access, its page is last in the algorithm's list, so its rank is $m$,
its wait is $\beta m$, and its cost is $(1+\beta)m$. Each block restores
the initial permutation and costs $(1+\beta)m^2$.

Keeping the initial list fixed and serving every request immediately is a
feasible offline schedule of cost $m(m+1)/2$ per block. Dividing the
algorithm's cost by this feasible upper bound on $\OPT$ gives the displayed
ratio. For any smaller proposed factor at a fixed $m$, repeating blocks
makes the resulting excess cost arbitrarily large, absorbing a fixed
additive constant. Finally let $m$ grow. This is a lower bound for the
specified threshold algorithm, not for all algorithms.
\end{proof}

\subsection{Why the nonlinear hypotheses matter}

\begin{proposition}[Concave powers]
\label{prop:concave-threshold-failure}
Let $f(t)=t^\gamma$, where $0<\gamma<1$. For every fixed $\beta>0$, the
rule that serves when its page's episode penalty reaches $\beta$ times
the current rank has unbounded competitive ratio, already on one page.
\end{proposition}
\begin{proof}
Put $\tau=\beta^{1/\gamma}$ and release $N$ requests at
$0,2\tau,\ldots,2\tau(N-1)$. The rule serves every request separately
at age $\tau$, before the following arrival. Its total cost is
$N(1+\beta)$. A feasible offline schedule serves all requests together
at the final arrival, for cost
\[
 1+[2\tau(N-1)]^\gamma.
\]
The ratio of these two costs grows as $\Omega(N^{1-\gamma})$, and the
ratio to $\OPT$ is at least as large. The excess over any proposed
constant multiple of the displayed comparison cost also diverges, so an
additive constant does not repair the guarantee. This disproves the
fixed-scale rule, not the existence of other algorithms for concave delay.
\end{proof}

\begin{proposition}[Request-dependent linear weights]
\label{prop:request-weights-failure}
Suppose requests to the same page may have different positive linear
weights, and the delay charge of an access is the maximum weighted waiting
time among the requests it clears. For every fixed $\beta>0$, the rule
that serves when this maximum reaches $\beta$ times the current rank has
unbounded competitive ratio, even on one page.
\end{proposition}
\begin{proof}
For $i=0,\ldots,N-1$, let request $i$ arrive at
\[
 a_i=1-4^{-i},\qquad w_i=2\beta\,4^i.
\]
Assuming earlier requests have been cleared, its weighted age reaches
$\beta$ at
\[
 s_i=a_i+\frac{1}{2\cdot4^i}
     <a_i+\frac{3}{4\cdot4^i}=a_{i+1}
\]
whenever a following request exists. Inductively, every request is served
alone and the algorithm pays $N(1+\beta)$.
Offline can instead serve the entire input at time one. Every weighted
age then equals
$w_i(1-a_i)=2\beta$, so this schedule costs only $1+2\beta$.
The ratio is unbounded as $N$ grows. Each instance is finite; the argument
does not require an infinite input accumulating at time one. There is no
conflict with a guarantee for a single fixed linear weight per page.
\end{proof}

\subsection{A limitation of averaging the global scale}

\begin{proposition}[The existing coefficient certificate]
\label{prop:random-scale-certificate}
Choose a positive random scale $B$ independently of the input and run the
fixed-scale rule with $\beta=B$. Averaging the deterministic inequalities
\[
 \ALG_\beta\le3(1+\beta)C_O+
                  2(1+\beta^{-1})D_O
\]
cannot certify a uniform competitive factor strictly below five. This is
only a limitation of these inequalities, not a randomized lower bound.
\end{proposition}
\begin{proof}
When the two moments are finite, the averaged spatial and delay
coefficients are $3(1+\E[B])$ and
$2(1+\E[B^{-1}])$. Both would have to be strictly below five for
this bound to certify a factor below five for arbitrary nonnegative
$C_O,D_O$. That would require
\[
 \E[B]<\frac23,
 \qquad \E[B^{-1}]<\frac32,
\]
contradicting
$\E[B]\,\E[B^{-1}]\ge1$ by Cauchy--Schwarz. If either
moment is infinite, the corresponding averaged coefficient gives no
finite improvement. The deterministic choice $B=2/3$ attains coefficient
five. A sharper analysis of randomized scales, or an algorithm that
resamples scales during its execution, is not ruled out by this argument.
\end{proof}

%% file: sections/disclosure.tex
\section*{Disclosure of AI-Assisted Research and Writing}
OpenAI language-model systems, including Codex, were used extensively in
developing this draft. Their contributions included proof exploration,
counterexample search, implementation and independent checking of small
instances, adversarial review of proofs, literature verification, and
LaTeX drafting and revision. The mathematical arguments are given in full;
computational checks are supplementary. Responsibility for the manuscript's
claims, citations, and final presentation remains with its authors.